\documentclass[twocolumn]{aastex631}

\usepackage{natbib}
\usepackage{amsmath}
\usepackage{xcolor}
\usepackage{hyperref}
\usepackage{academicons}

\newcommand{\Kepler}{{\sl Kepler}\ }
\newcommand{\Keplers}{{\sl Kepler's} }

\shorttitle{Implicit biases from pseudo-density}
\shortauthors{Gilbert, MacDougall, \& Petigura}

\graphicspath{{./}{Figures/}}

\begin{document}

\title{Implicit biases in transit models using stellar pseudo-density}

\correspondingauthor{Gregory J. Gilbert}
\email{gjgilbert@astro.ucla.edu}

\author{Gregory J. Gilbert}
\affiliation{Department of Physics and Astronomy, University of California, Los Angeles}

\author{Mason G. MacDougall}
\affiliation{Department of Physics and Astronomy, University of California, Los Angeles}

\author{Erik A. Petigura}
\affiliation{Department of Physics and Astronomy, University of California, Los Angeles}

\begin{abstract}

The transit technique is responsible for the majority of exoplanet discoveries to date. Characterizing these planets involves careful modeling of their transit profiles. A common technique involves expressing the transit duration using a density-like parameter, $\tilde{\rho}$, often called the ``circular density.'' Most notably, the \Kepler project -- the largest analysis of transit lightcurves to date -- adopted a linear prior on $\tilde{\rho}$. Here, we show that such a prior biases measurements of impact parameter, $b$, due to the non-linear relationship between $\tilde{\rho}$ and transit duration. This bias slightly favors low values ($b \lesssim 0.3$) and strongly disfavors high values ($b \gtrsim 0.7$) unless transit signal-to-noise ratio is sufficient to provide an independent constraint on $b$, a criterion that is not satisfied for the majority of \Kepler planets. Planet-to-star radius ratio, $r$, is also biased due to $r{-}b$ covariance. Consequently, the median \Kepler DR25 target suffers a $1.6\%$ systematic underestimate of $r$. We present a techniques for correcting these biases and for avoiding them in the first place.

\end{abstract}


\section{Introduction}\label{sec:intro}

In the two decades since the discovery of the first transiting hot Jupiter \citep{Charbonneau2000, Henry2000}, the transit technique has grown to be the most prolific exoplanet detection method to date, accounting for 77\% of the current census. Contemporary work continues to rely heavily on the transit technique. To wit, several transit-focused NASA and ESA missions are either already on-sky (\textit{TESS}, \citealt{Ricker2015}) or slated for launch in the near future (\textit{PLATO}, \citealt{Rauer2014}), and next-generation radial velocity spectrographs have been designed for follow-up characterization of known transiting planets (e.g. KPF, \citealt{Gibson2016}; MAROON-X, \citealt{Seifahrt2018}). The transit technique will remain indispensable for exoplanet astronomy for decades to come.

Accurate modeling of the transit lightcurve is a critical step for characterizing transiting planets. At the most basic level, transit modeling involves computing the time-dependent flux $F(t)$ of a star obscured by a transiting planet relative to the unobscured flux $F_0$. If one assumes a spherical planet and star, this computation depends strictly on the planet-to-star size ratio $r$, the (time-dependent) center-to-center sky-projected planet-to-star separation $z$ (measured in units of $R_{\star}$), and the radial dependence of the stellar limb-darkening profile $\{u\}$. Early analyses computed $F(z;r,\{u\})$ via numerical integration, but today the most widely used method is the \citet{MandelAgol2002} model, which expresses the transit lightcurve via an analytic solution to $F(z;r,\{u\})$ for several limb darkening profiles which can be described by a small set of limb-darkening parameters.

In order to model time-series photometry, one must convert $F(z;r,\{u\})$ into $F(t;r,\{u\})$. While $z$ is the only parameter that varies with time, one may choose how to specify the function that maps $t \rightarrow z$. If one assumes strict periodicity of transits and a constant projected velocity during transit., then in the limit $r \rightarrow 0$, $z(t)$ may be specified completely by an orbital period, $P$, a transit mid-point, $t_0$, an impact parameter, $b$, and 1st-to-4th contact transit duration, $T_{14}$.\footnote{Several alternative transit durations may be substituted for $T_{14}$: (1) the 2nd-to-3rd contact duration, $T_{23}$, (2) the center-to-center contact duration, $T_{\rm cc}$, also called the 1.5-to-3.5 contact duration, or (3) the full-width-half-max duration, $T_{\rm FWHM}$, which may be defined in relation to the transit depth. While each has its merits (see \citealt{Kipping2010-duration} for discussion), we adopt $T_{14}$ throughout this work because it is the transit duration which is most readily defined for all grazing and non-grazing transit geometries.} This parameterization --- $F(t; P, t_0, r, b, T_{14})$ --- is convenient and is closely linked to the transit geometry.

An alternative approach is to specify $T_{14}$ from a combination of scaled separation $a/R_{\star}$, orbital eccentricity $e$, argument of periastron $\omega$, and projected inclination $\cos i$, following \cite[][]{Winn2010} as

\begin{equation}\label{eq:t14_winn2010}
    T_{14} \simeq  \frac{P}{\pi}\sin^{-1}\Big(\frac{R_{\star}}{a}\frac{\sqrt{(1+r)^2 - b^2}}{\sin i}\Big)
    \Big(\frac{\sqrt{1-e^2}}{1+e\sin\omega}\Big)
\end{equation}

\begin{equation}
    b = \frac{a\cos i}{R_{\star}}\Big(\frac{1-e^2}{1+e\sin\omega}\Big)
\end{equation}

\noindent Now the lightcurve is specified by the function $F(t; P, t_0, r, a/R_{\star}, b, e, \omega)$, which is similar to the parameterization used by the \texttt{EXOFAST} suite \citep{Eastman2013, Eastman2017}.\footnote{In practice, \texttt{EXOFAST} uses $\log(a/R_{\star})$ and expresses $b$ as $\cos i$; $\{e,\omega\}$ is usually specified as $\{\sqrt{e}\sin\omega,\sqrt{e}\cos\omega\}$ in order to establish uniform priors on $e$ and $\omega$ and to a avoid a boundary issue at $e=0$.} A related approach is to replace $a/R_{\star}$ with stellar density by employing Kepler's third law. Thus, the light curve may also be parameterized by $F(t; P, t_0, r, \rho_{\star}, \cos i, e, \omega)$.

These two eccentricity-explicit parameterizations have the advantage that the lightcurve has been specified completely by properties of the star, planet, and planetary orbit; the disadvantage is that five parameters have been replaced by seven, and thus significant degeneracies between $\{a/R_{\star}, e, \omega\}$ or $\{\rho_{\star}, e ,\omega\}$ are inevitable. These degeneracies lead to inefficiencies with light curve fitting and posterior sampling.

A common shortcut is to fit the lightcurve assuming that $e = 0$ even though the orbit may, in fact, be eccentric. This assumption reduces the number of free parameters back to five, but $\rho_{\star}$ can no longer be thought of as a stellar density. Rather, it is a stand-in for duration which merely has \textit{units} of density, defined by \citet{SeagerMallenOrnelas2003} as
\begin{equation}\label{eq:rho_smo}
    \tilde{\rho} \equiv \Bigg(\frac{4\pi^2}{P^2G}\Bigg) 
    \Bigg(\frac{(1+r)^2 - b^2\big(1 - \sin^2[\pi T_{14}/P]\big)}{\sin^2[\pi T_{14}/P]}\Bigg)^{3/2}
\end{equation}
where $G$ is Newton's gravitational constant. This quantity $\tilde{\rho}$ is sometimes referred to as the ``mean stellar density,'' the ``circular density,'' or the ``observed density,'' but we prefer to call it the ``pseudo-density'' because (1) the other names are confusing, and (2) $\tilde{\rho}$ matches the true stellar density only when numerous assumptions are met \citep[see][]{Kipping2014-asterodensity}.

Because the prior expectation for $\tilde{\rho}$ is a complicated function of $\rho_{\star}$, $b$, $e$, and $\omega$, na{\"i}vely placing a flat prior on $\tilde{\rho}$ and adopting it as a fitting parameter induces undesired biases on $T_{14}$ and $b$.

To date, $\tilde{\rho}$ has enjoyed widespread use in the exoplanet literature. For example, the \Kepler project (\citealt{Borucki2010}; the largest analysis of transit lightcurves to date) fit their lightcurves with the $F(t; P, t_0, r, \tilde{\rho}, b)$ parameterization \citep{Rowe2014, Rowe2015, Mullally2015, Coughlin2016, Thompson2018}.  We discuss the effects of that choice in \S\ref{sec:kepler}. More broadly, this paper investigates the implicit biases on impact parameter and other light curve parameters that result from the use of $\tilde{\rho}$. 

Throughout this work, we assume that all transit signals under investigation have been thoroughly vetted such that the detected signal is known to be a real transit at high confidence. The methods employed in this work are thus appropriate for parameter estimation but \underline{not} for transit detection or vetting.

This paper is organized as follows. In \S\ref{sec:motivation} we empirically demonstrate the origin of the $\tilde{\rho}$ bias by fitting a transit lightcurve model to simulated photometry using the \Kepler project parameterization; we then demonstrate that our preferred parameterization does not suffer from this bias. In \S\ref{sec:experiment} we present a numerical experiment which isolates the effects of various model assumptions on posterior inferences. In \S\ref{sec:jacobian} we analytically derive the Jacobian of the coordinate transformation $T_{14} \rightarrow \tilde{\rho}$ which explains the origin of the empirical bias. In \S\ref{sec:kepler} we show that the $\tilde{\rho}$ bias has affected most posterior inferences of $b$ and $r$ derived from \Kepler data. In \S\ref{sec:conclusion} we summarize our conclusions and discuss other biases which arise from using related parameterizations such as $a/R_{\star}$.

\section{Understanding parameter biases with fits to synthetic photometry}\label{sec:motivation}

\begin{figure}
    \centering
    \includegraphics[width=0.45\textwidth]{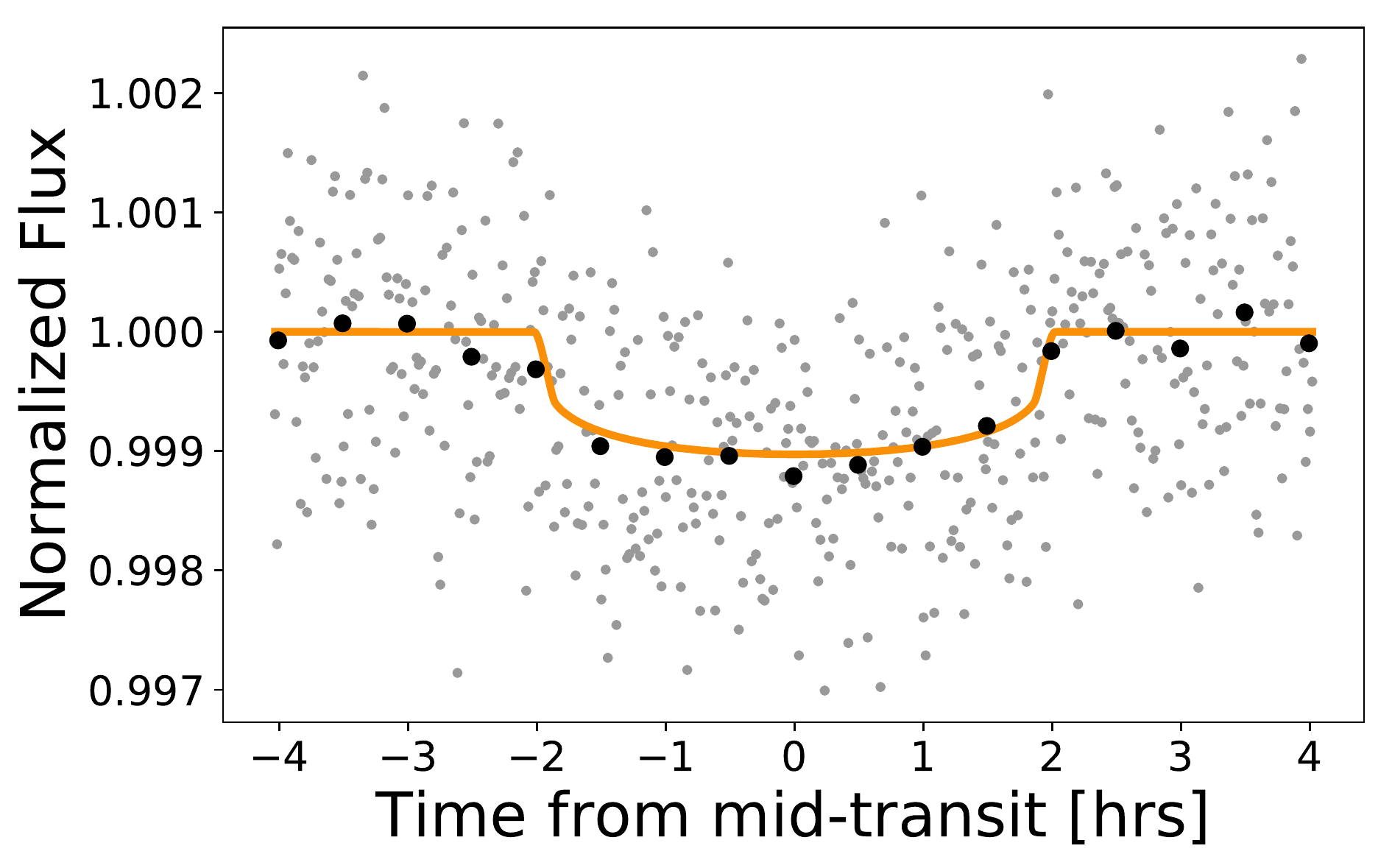}
    \caption{Simulated photometry for a mini-Neptune on a circular 15 day orbit around a Sun-like star, transiting at $b=0.5$. The orange line indicates the ground truth transit model. Grey points show simulated observations at a one minute observing cadence; black circles are binned to 30 minutes. The white noise level was set to S/N = 16, close to the \Kepler median. See Table \ref{tab:sim_params} for ground-truth simulation parameters.}
    \label{fig:sim_photometry}
\end{figure}

To illustrate the $\tilde{\rho}$ bias, we simulated photometric observations of a warm mini-Neptune ($P=15$ days, $r_p = 3.3\ R_{\oplus}$) on a circular orbit around a Sun-like star, transiting at impact parameter $b=0.5$. We simulated data with a 30 minute observing cadence (matching \Keplers long cadence observing mode) within $\pm T$ from the transit center. All photometric data were oversampled by a factor of 7 and integrated using Simpson's rule to account for the effects of finite integration time \citep{Kipping2010-binning}. The white noise level was tuned to produce S/N = 16, which is slightly lower than the median \Kepler value and results in a posterior model with $\sigma_r/r \approx 0.10$ and $\sigma_T/T \approx 0.05$, where $\sigma_r/r$ corresponds to the fractional posterior measurement, and similar for $T$. We chose these values in order to produce a transit which is similar to those found by \textit{Kepler}. Ground-truth simulation parameters are listed in Table \ref{tab:sim_params}, and simulated photometry is shown in Figure \ref{fig:sim_photometry}.

\begin{deluxetable}{ll}
\tablecaption{Ground-truth simulation parameters; simulated photometry is shown in Figure \ref{fig:sim_photometry}.}
\label{tab:sim_params}
\tabletypesize{\scriptsize}
\tablehead{
\colhead{Parameter} & \colhead{Value} \\ 
} 
\startdata
$M_{\star} [M_{\odot}]$ & $1.0$ \\
         $R_{\star} [R_{\odot}]$ & $1.0$ \\
         $u_1, u_2$ & $0.40, 0.25$ \\
         $P$ [days] & $15.0$ \\
         $r$ & $0.03$ \\
         $b$ & $0.5$ \\
         $T_{14}$ [hrs] & $3.29$ \\
         $S/N$ & $16$ \\
\enddata
\end{deluxetable}

The transit model was specified using a standard pseudo-density parameterization: $\{P, t_0, r, b, \tilde{\rho}\}$. In order to minimize confounding factors, we held $P$ and $t_0$ fixed at their injected values; we also held the mean out-of-transit flux, $F_0$, and photometric white noise level, $\sigma^2_{\rm phot}$, fixed to their true values, which is equivalent to assuming the raw photometry has been accurately pre-whitened. For the remaining transit parameters, we adopted broad weakly informative priors with permissive bounds (see Table \ref{tab:priors} for details), for a total of three free parameters per model: $\{r, b, \tilde{\rho}\}$ (the ``$\tilde{\rho}$ basis''), or $\{\log r, b, \log T_{14}\}$ (the ``$\log T$ basis''). We chose the later basis because $T_{14}$ is typically well constrained by the data and furthermore may be assigned priors in a sensible fashion; sampling in $\log r$ and $\log T_{14}$ is equivalent to placing log-uniform priors on $r$ and $T_{14}$ which facilitates exploration of posterior values over different orders of magnitude. We modeled a circular transit in all cases and held stellar mass, radius, and limb darkening to their true values during the fit; there is no loss of generality in this approach, because as long as we ignore minuscule ingress/egress asymmetry that exists for eccentric transits \citep{Barnes2007}, there is no difference between a circular and eccentric transit. In order to avoid complications which arise when modeling grazing transits, we restricted impact parameters to $b < 1-r$.\footnote{A common approach (which we did not adopt) is to draw samples uniformly from the $r-b$ plane using triangular sampling \citep{Espinoza2018}. However, naive application of this method induces a marginal prior on $r$, so caution must be taken to ensure that priors are established as intended.} To confirm that this restriction is permissible, we explored the parameter space near the limb of the star following the methodology of \citet{Gilbert2022} and verified that the simulated transit is inconsistent with a grazing geometry.

\begin{deluxetable}{ll | ll}
\tablecaption{Priors on model parameters for simulated lightcurve.}
\label{tab:priors}
\tabletypesize{\scriptsize}
\tablehead{
\colhead{Parameter} & \colhead{Value} & \colhead{Parameter} & \colhead{Value} \\ 
} 
\startdata
         $r$ & $\mathcal{U}(0.01,0.1)$ & $\log r$ & $\mathcal{U}(-2, -1)$ \\
         $b$ & $\mathcal{U}(0,1-r)$ & $b$ & $\mathcal{U}(0,1-r)$ \\
         $\tilde{\rho}/\rho_{\odot}$ & $\mathcal{U}(0.1,10)$ & $\log [T_{14}/ \rm hr]$ & $\mathcal{U}(1,10)$ \\
\enddata
\end{deluxetable}

We drew samples from the posterior using Hamiltonian Monte Carlo \citep[HMC;][]{Neal2011} and the No U-Turn Sampler \citep[NUTS;][]{Hoffman2011}. Each model iteration consisted of two chains run for 5000 tuning steps and 20,000 draws, producing an effective number of samples greater than 11,000 for all parameters for each of the the two parameterizations. 

\begin{figure}
    \centering
    \includegraphics[width=0.45\textwidth]{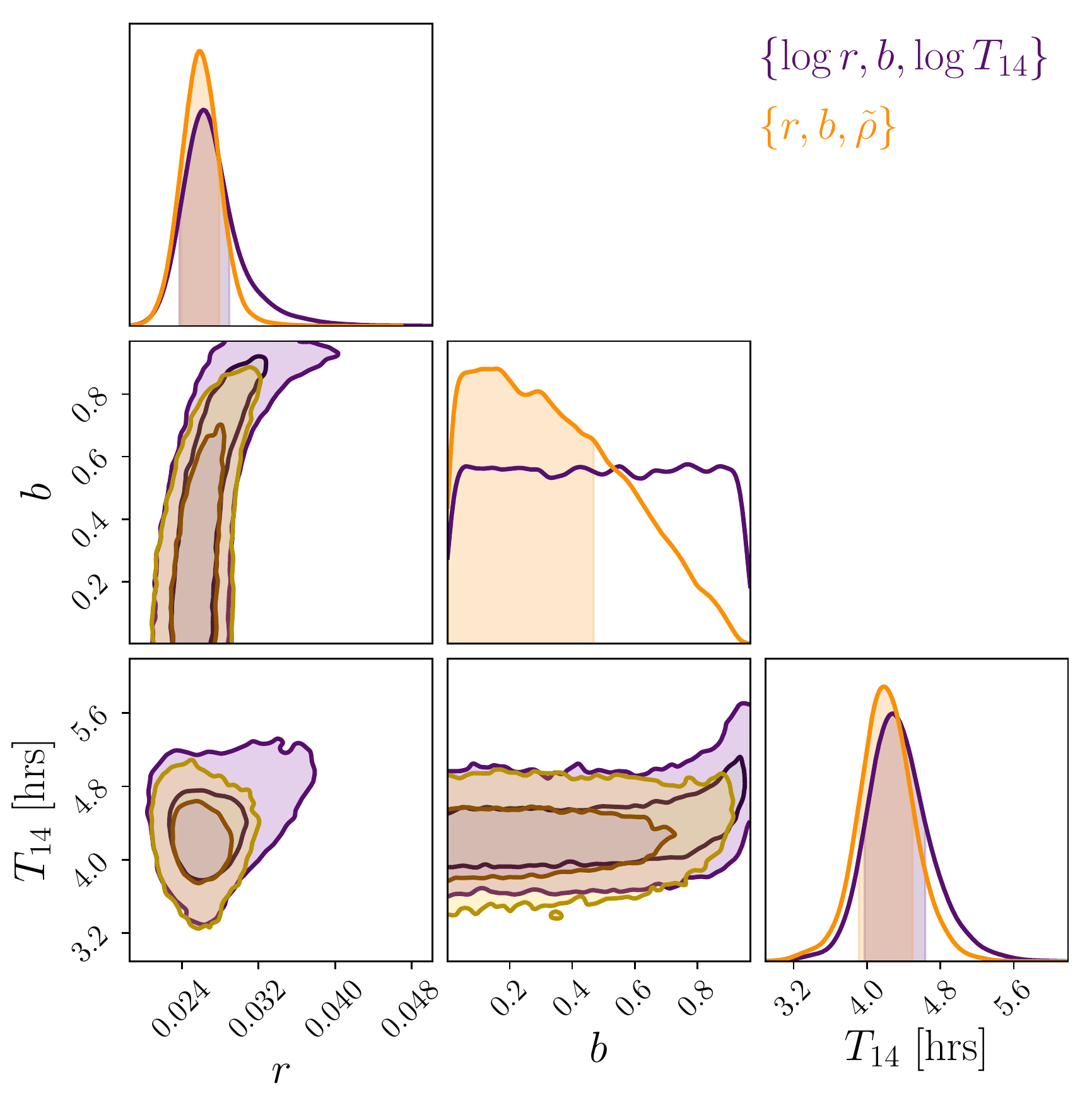}
    \caption{Posterior corner plots when sampling in the $\tilde{\rho}$ basis (orange) vs the $\log T_{14}$ basis (purple). The bias on impact parameter, $b$, is apparent when sampling with $\tilde{\rho}$ but is resolved when sampling in $T_{14}$.}
    \label{fig:corner_overlay}
\end{figure}

Posterior corner plots for the quantities of interest are shown in Figure \ref{fig:corner_overlay}. The most notable difference is in the 1D marginalized distribution of impact parameter. When sampling using the $\tilde{\rho}$ basis, the posterior is biased toward low $b$; as a point of reference, 74\% of the probability mass is below $b = 0.5$, the injected value. When sampling using the $\log T$ basis, however, the distribution of $b$ is nearly uniform over the allowed range, reflecting the fact that for a low signal-to-noise transit the impact parameter is largely unconstrained.  Our results did not substantially change when simulating a one minute observing cadence (matching \Keplers short cadence mode), indicating that the $\tilde{\rho}$ bias arises from the model parameterization and is not an artifact of data binning. We also repeated the analysis using $r=0.1$ and $r=0.01$ and found that the results did not change.

Clearly, the results are inconsistent between models -- which contain identical underlying physics and differ only in their parameter bases -- so at least one of the two models has produced biased inference. In the sections that follow, we present both a numerical argument (\S\ref{sec:experiment}) and an analytic argument (\S\ref{sec:jacobian}) which demonstrate that the $\log T$ basis has produced the desired result.

\section{Numerical sampling experiment}\label{sec:experiment}

We will now demonstrate that the bias on $b$ seen in the previous section arises solely from the model parameterization and not from vagaries of the MCMC sampling algorithm or peculiarities of the noise realization in the photometry.

To do so, we performed a numerical experiment which approximated the lightcurve modeling procedure from \S\ref{sec:motivation} by drawing samples directly from the prior distributions and then applying an \textit{a posteriori} importance weighting designed to mimic the constraints imposed by the photometry. When determining these importance weights, we employed a Gaussian likelihood function and approximated the (covariant) parameter constraints from \S\ref{sec:motivation} as independent univariate Gaussians. The key advantage of this method is that we no longer needed to directly fit the photometry, thereby eliminating potential confounding factors introduced by the photometry and the sampler.

\subsection{Experimental setup}\label{subsec:experiment_setup}

We adopted the same fiducial star-planet system as \S\ref{sec:motivation}, placing a $3.3 R_{\oplus}$ mini-Neptune on a circular 15 day orbit around a solar twin. We fixed the ephemeris $\{P, t_0\}$ throughout and placed uniform interval priors on all other parameters $\{r, b, \tilde{\rho}, \log T_{14}\}$ as before (see Table \ref{tab:priors}), with the small modification that we now allow $b$ to range over all detectable values, i.e. $b \sim \mathcal{U}(0,1+r)$; this modification is acceptable because our sampling procedure (see below) avoids the usual issues which arise when fitting grazing transits \citep[see][]{Gilbert2022}.

For the first iteration of the experiment we adopted the $\tilde{\rho}$ basis $\{r, b, \tilde{\rho}\}$ and drew random samples directly from the prior distributions. We next calculated transit duration using
\begin{equation}\label{eq:T14_approx}
    T_{14} = \frac{PR_{\star}}{\pi a}\Big((1+r)^2 - b^2\Big)^{1/2}
\end{equation}
for each sample. Here, we have approximated Equation \ref{eq:t14_winn2010} by using the small angle approximation $\sin^{-1}\phi \approx \phi$ and $i \approx \pi/2 \rightarrow \sin i \approx 1$. The scaled separation can be calculated from Kepler's Third Law as $a/R_{\star} = [(GP^2\rho)/(3\pi)]^{1/3}$. 

For subsequent iterations of the experiment, we modified the procedure to use use three alternative parameter bases: (1) $\{\log r, b, \log{\tilde{\rho}}\}$, (2) $\{r, b, T_{14}\}$, and (3) $\{\log r, b, \log T_{14}\}$. We chose these parameterizations in order to explore the effects of uniform vs log-uniform priors in addition to the effect of substituting $\tilde{\rho} \rightarrow T_{14}$. We followed the same sampling procedure as before, except when drawing samples of $T_{14}$ or $\log T_{14}$ we calculated  $\tilde{\rho}$ following Equation \ref{eq:rho_smo}.

Mimicking the simulated light curve in \S2, we assumed that we could constrain $r$ to $10\%$ accuracy and $T_{14}$ to $5\%$ accuracy, with independent Gaussian precision from the photometry (i.e. $\sigma_r/r = 0.1$, $\sigma_T/T = 0.05$). We further assumed that the impact parameter would be entirely unconstrained by the data. These uncertainties are representative of typical values, but we have removed the covariance and forced them to be Gaussian (or unconstrained), which eases interpretation.

We imposed our assumed measurement uncertainties on $r$ and $T_{14}$  by calculating the log-likelihood of each $i^{\rm th}$ sample
\begin{equation}\label{log-like}
    \log \mathcal{L}_i = -\frac{1}{2}\Big(\frac{T_i - T_{\rm true}}{\sigma_T}\Big)^2 
    -\frac{1}{2}\Big(\frac{r_i - r_{\rm true}}{\sigma_r}\Big)^2
\end{equation}
which assumes a Gaussian likelihood function. We then weighted each sample by 
\begin{equation}
    w_i = \frac{\mathcal{L}_i}{\sum_i \mathcal{L}_i}
\end{equation}
to produce our synthetic posterior distributions.

\subsection{Bias on impact parameter}

\begin{figure*}
    \centering
    \includegraphics[width=0.95\textwidth]{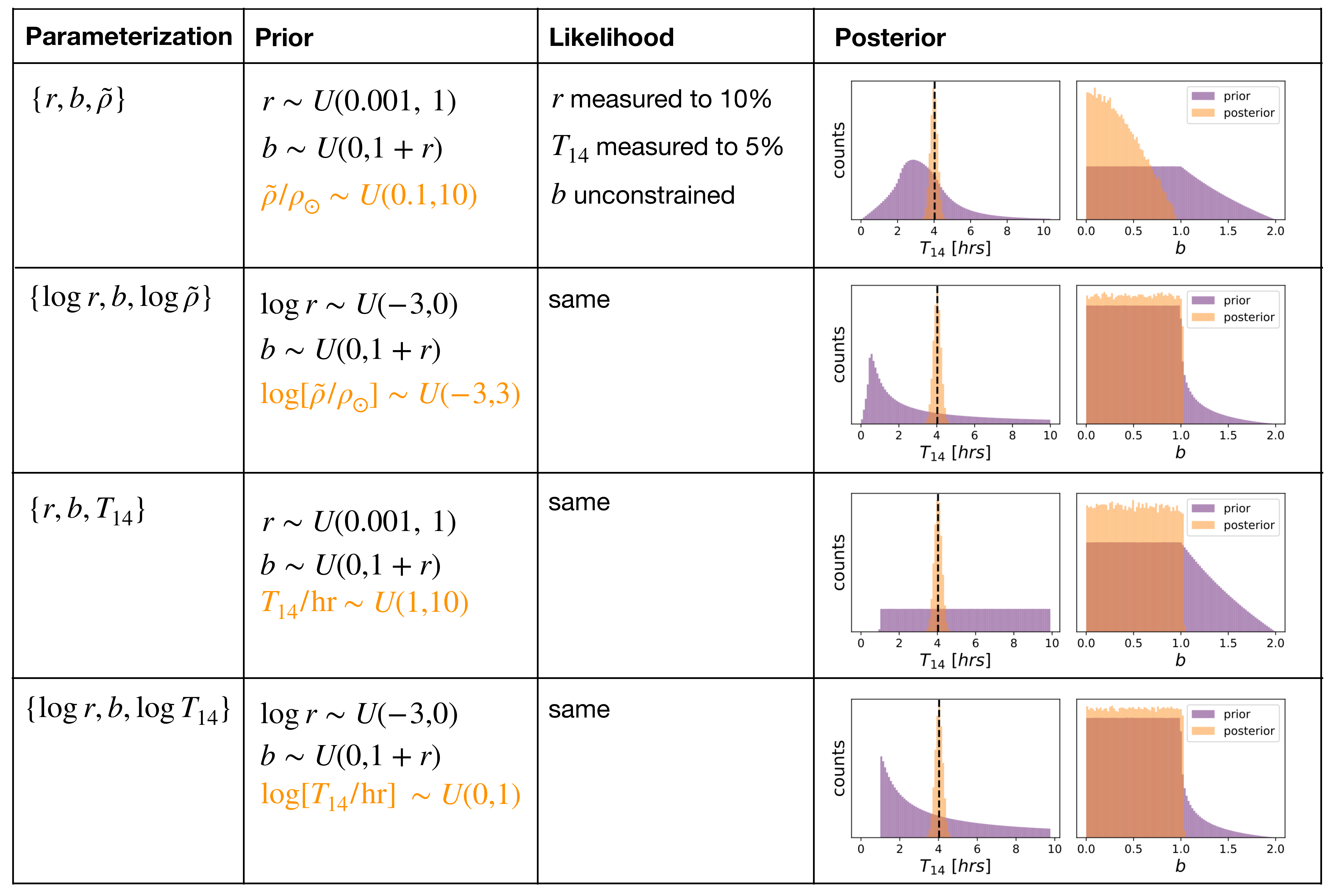}
    \caption{Results of the numerical sampling experiment described in \S\ref{sec:experiment}. Each row corresponds to the prior, likelihood, and posterior for a given model parameterization. For visual clarity, the height of the $T_{14}$ posterior has been reduced by a factor of 3 on all plots. The difference in the prior distribution on $b$ for rows 1 \& 3 compared to rows 2 \& 4 stems from the use of $r$ vs $\log r$, respectively. Sampling with a uniform prior on $\tilde{\rho}$ (top row) produces a nonuniform prior on $T_{14}$ and a biased posterior for $b$. In contrast, sampling in any of the other parameter bases produces a posterior estimate of $b$ which matches the prior, except in cases where constraints on $r$ would produce a non-transiting orbit.}
    \label{fig:impact_bias}
\end{figure*}

The results of our numerical experiment are summarized in Figure \ref{fig:impact_bias}. As expected, when parameterizing the model as $\{r, b, \tilde{\rho}\}$ with uniform priors, we obtain biased results that are qualitatively similar to those produced in \S\ref{sec:motivation} (i.e. by fitting the photometry directly). Notably, sampling in $\tilde{\rho}$ produces a strong prior on $T_{14}$ (purple) which is not physically motivated. Because $T_{14}$ is constrained to 5\%, the data overwhelm the prior and the $T_{14}$ posterior distribution (orange) is only slightly biased. The posterior on impact parameter, however, is clearly different from the prior even though our model included no information about impact parameter. Because we have (by construction) placed no measurement constraint on $b$, the posterior distribution should match the prior. In reality however, the posterior is tilted toward $b=0$, giving the illusion of a (modestly) constrained posterior. 

The $\tilde{\rho}{-}b$ bias is resolved by using any of the alternative parameterizations which substitute $\log\tilde{\rho}$, $T_{14}$, or $\log T_{14}$ for $\tilde{\rho}$. Although using the substitution $\tilde{\rho} \rightarrow \log\tilde{\rho}$ may seem at first glace to be the simplest choice (requiring little change from existing practices), we argue that using either of the duration-based parameterizations is preferable for two reasons. First, the results are insensitive to the exact choice of (reasonable) prior placed on $T_{14}$, whereas they are highly sensitive to the prior placed on $\tilde{\rho}$; insensitivity to priors is in general a desirable feature of robust inference. Second, setting prior interval bounds on $\tilde{\rho}$ is a non-intuitive task, requiring careful consideration of the true stellar density and orbital elements. In contrast, principled priors may be placed on the transit duration quite simply following inspection of the transit lightcurve. In fact, setting bounds on $T_{14}$ is so straightforward that it could even be done algorithmically following the output of a box-least squares transit search \citep{Kovacs2002}. The bottom line is that given the choice between options which produce equivalent results, we prefer the simpler of the two.

In summary, because we have decoupled the posteriors from complicating factors (e.g. parameter covariances, sampler inefficiencies, etc.), we conclude that the differences between posterior distributions obtained under the $\tilde{\rho}$ basis versus the $\log T_{14}$ basis arise solely due the parameterization. Furthermore, we conclude that the $\tilde{\rho}$ basis (with a uniform prior) induces a bias on $b$, whereas the other options we have presented produce unbiased estimates.

\section{Mathematical origin of the bias}\label{sec:jacobian}

In the previous sections, we illustrated the biases on $b$ that result from uniform and log-uniform priors on $\tilde{\rho}$ by exploring synthetic photometry fits and simple numerical experiments. In this section, we investigate the mathematical origins of this bias.

The transit parameter covariance matrix was previously derived by \citet{Carter2008}, but where their treatment prioritized analytic interpretability (with a small sacrifice to accuracy), our treatment prioritizes accuracy (with a small sacrifice to interpretability). Most importantly, the covariance matrix derived by \citet{Carter2008} are least accurate as $b \rightarrow 1$ and in the presence of non-neglible limb darkening, which are precisely the conditions under which the $\tilde{\rho}$ bias we are investigating become most important. Thus, our work complements rather than supplants \citet{Carter2008}.

When modeling light curves, our main goal is to derive the posterior probability density function, $p(\vec{x})$, i.e. the probability that a set of planet properties $\vec{x}$ resides in an infinitesimal volume element spanning $\vec{x}$ to $\vec{x} + d\vec{x}$. However, this probability is not invariant under changes in parameterization.  Specifically, for our problem, $p(T_{14})/dT_{14} \neq p(\tilde{\rho})/d\tilde{\rho}$. To convert $p(T_{14})$ to $p(\tilde{\rho})$, one must account for the change in infinitesimal volume element resulting from the $T_{14} \rightarrow \tilde{\rho}$ transformation, i.e. the Jacobian
\begin{equation}\label{eq:approx_jacobian}
    J = \frac{d\tilde{\rho}}{dT_{14}} = 
     -\frac{12\pi^3}{P^3G}
    \Bigg((1+r)^2 - b^2\Bigg)^{3/2}
    \Bigg(\frac{\pi T_{14}}{P}\Bigg)^{-4}
\end{equation}
which we derive in the Appendix. The Jacobian of the transformation $T_{14} \rightarrow \log\tilde{\rho}$ is simply
\begin{equation}
    J' = \frac{d\log\tilde{\rho}}{dT_{14}} = -\frac{3}{T_{14}}
\end{equation}
which is independent of $b$, explaining why using $\log\tilde{\rho}$ in place of $\tilde{\rho}$ produces unbiased posteriors.

\begin{figure}
    \centering
    \includegraphics[width=0.45\textwidth]{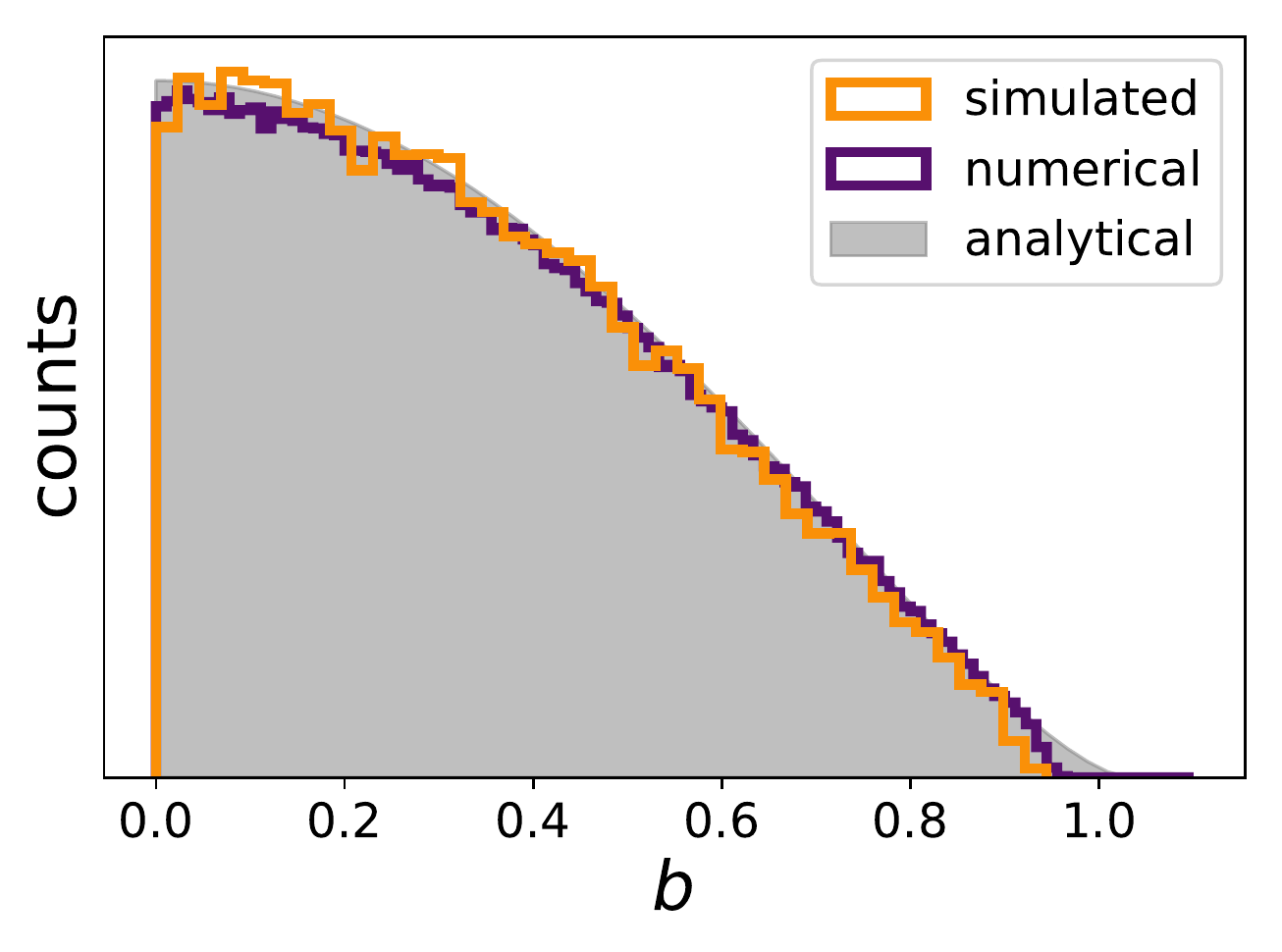}
    \caption{Posterior samples of $b$ from the simulated transit fit (orange histogram, \S\ref{sec:motivation}) and the numerical experiment (purple histogram, \S\ref{sec:experiment}) are nearly perfectly matched by the expected bias from the analytically derived Jacobian (grey shaded region, \S\ref{sec:jacobian}).}
    \label{fig:analytic_b_bias}
\end{figure}

In Figure \ref{fig:analytic_b_bias}, we show the analytic Jacobian in Equation \ref{eq:approx_jacobian} alongside the simulated posterior samples of $b$ obtained in \S\ref{sec:motivation} and the numerical results obtained in \S\ref{sec:experiment}. It is evident from inspection that the distributions are in close agreement. We conclude that the non-uniform distribution of $b$ arises from the combination of parameterization and (incorrect) prior, rather than from any real constraint imposed by the data.

\section{Biased Kepler planet properties}\label{sec:kepler}

\begin{figure*}
    \centering
    \includegraphics[width=0.95\textwidth]{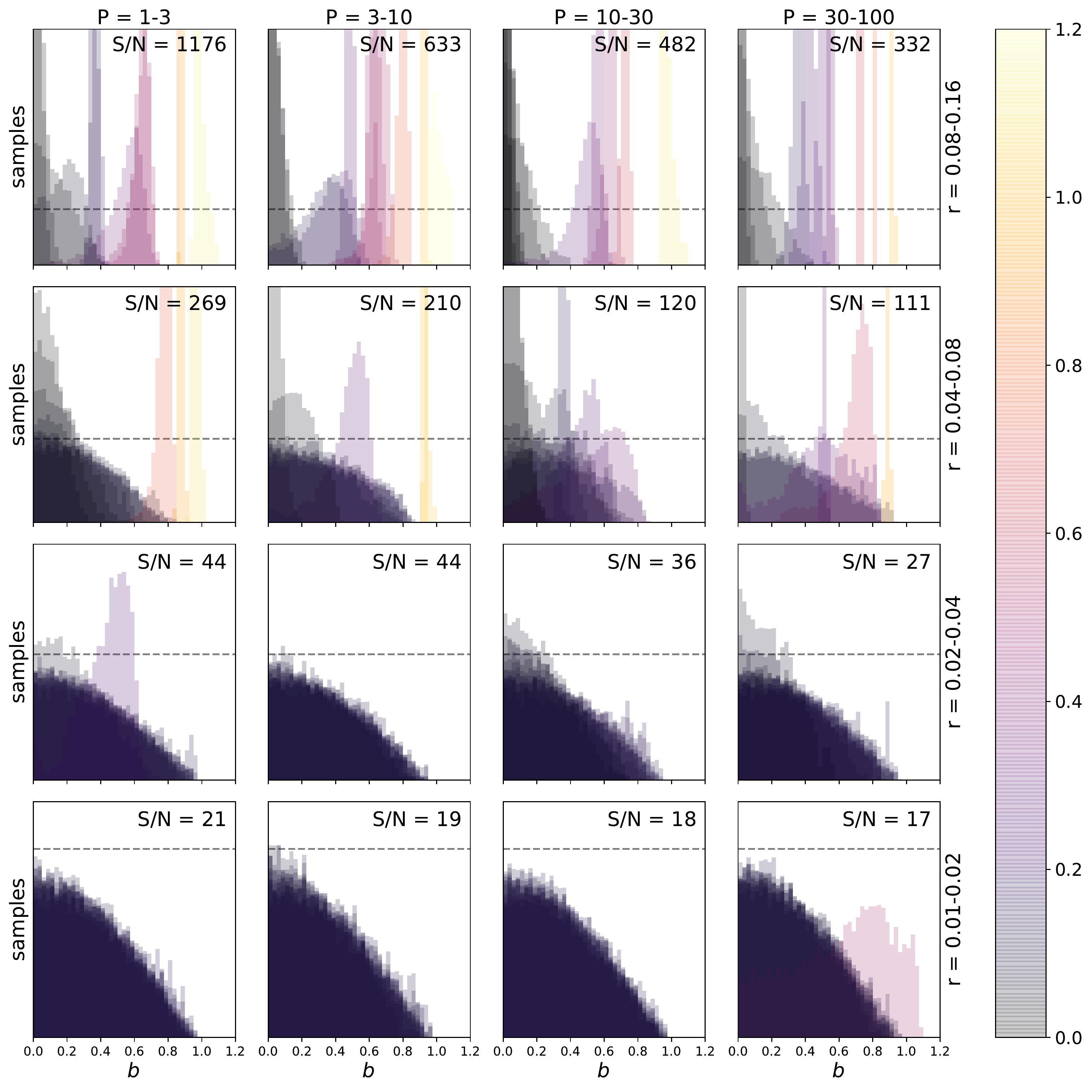}
    \caption{Posterior distributions of impact parameter for a random selection of KOIs, organized in logarithmic bins on a $P-r$ grid. Data shown are the posterior MCMC chains from \Kepler Data Release 25 \citep{Thompson2018}, described in detail in \citet{Rowe2014} and downloaded from the NASA Exoplanet Archive \citep{Akeson2013}. Each posterior distribution is plotted with 20\% opacity so that dark regions indicate where many distributions overlap; colors correspond to the median $b$ value for a given KOI. For visual clarity, a maximum of twelve KOIs are plotted per panel. The horizontal axis of each panel ranges over $b=(0,1.2)$; the vertical range of each row is different, but the dashed line indicates the same distribution height. The median S/N in each 2D bin is printed in the upper right-hand corner of the grid squares. It is clear from inspection that most of the objects (excluding the largest, highest S/N objects) show qualitatively similar posterior distributions of $b$. The similarity is particularly striking for small (low S/N) objects.}
    \label{fig:kepler_gallery}
\end{figure*}

We have shown that adopting a linear $\tilde{\rho}$ prior results in a biased impact parameter. The \Kepler project \citep{Borucki2010, Rowe2014, Rowe2015, Mullally2015, Coughlin2016, Thompson2018} used such a parameterization (Jason Rowe, private communication). Therefore, we expect biased $b$ in all cases except those where $b$ is strongly constrained by the light curve itself. Because most \Kepler planet candidates exhibit modest transit signal-to-noise (median S/N = 22.4), the characteristic ``hill'' shape we have seen for biased posterior $b$ distributions in the previous three sections is also present in the posterior distributions of nearly every \Kepler planet candidate from DR25 \citep{Thompson2018}. Figure \ref{fig:kepler_gallery} illustrates the presence of the $b$ bias over a grid of orbital periods and radii. Only the largest (and therefore highest S/N) planets consistently exhibit meaningful constraints on $b$.

Due to signal-to-noise bias which disfavors the detection of high-$b$ transits \citep{KippingSandford2016}, the prior expectation on impact parameter is not exactly flat, and so the posteriors exhibited in \Kepler data will not exactly match the idealized distribution we derived in \S\ref{sec:motivation}-\ref{sec:jacobian}. However, most \Kepler detections have $S/N > 10$ and fall in the flat part of the detection completeness curve \citep{Christiansen2020}. Thus, the appropriate prior for the vast majority of \Kepler planets should be nearly flat in $b$, with a fall off at the value of $b$ that reduces $S/N$ to $\sim\ 10$.

\begin{figure*}
    \centering
    \includegraphics[width=0.85\textwidth]{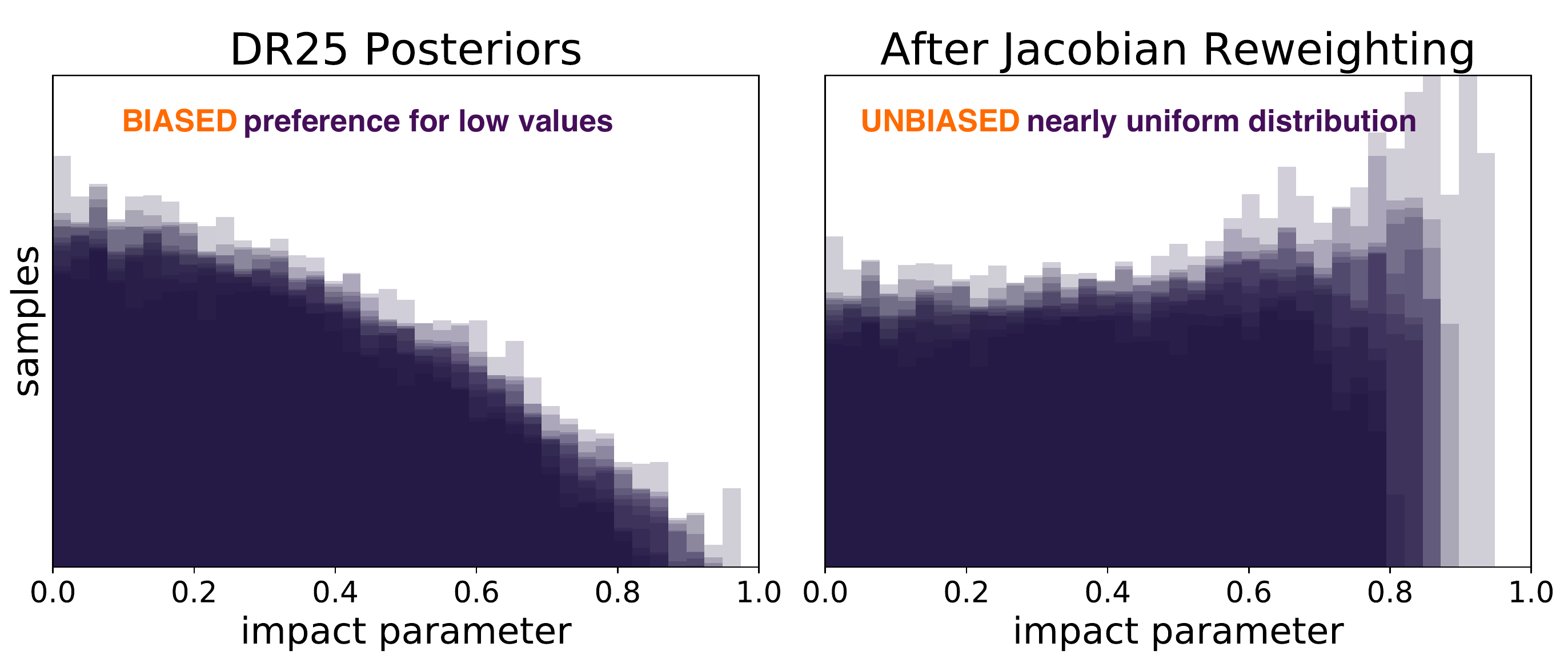}
    \caption{Posterior samples of $b$ from a representative selection of DR25 targets before and after reweighting by the Jacobian to correct for biases induced by sampling in $\tilde{\rho}$. All targets have $0.02 < r < 0.04$ and $10 < P < 30$ days. \textit{Left panel}: raw DR25 posteriors chains show a clear (biased) preference for low values of $b$. \textit{Right panel}: after reweighting, the (unbiased) distribution is nearly flat. To minimize spurious peaks and sampling noise in low probability regions, the lowest density 1\% of samples have been excluded from our reweighting scheme. The slight increase in probability density near $b\approx1$ in the reweighted posteriors reflects the presence of residual importance sampling noise rather than a real feature of the data. Because there is significant sampling noise (due to the large implied posterior mass in regions with few samples), our preferred method for ameliorating the pseudo-density bias is to refit the photometry.}
    \label{fig:b_posteriors_before_and_after}
\end{figure*}

Detection biases notwithstanding, the $\tilde{\rho}$ bias is easily understood and corrected. Because the relationship between $\tilde{\rho}$, $b$, $r$, and $T_{14}$ is known analytically \citep{SeagerMallenOrnelas2003}, one needs only to apply the appropriate Jacobian weighting in order to transform an unintended prior on $\tilde{\rho}$ into the desired prior on $b$ or $T_{14}$ (or any other basis parameter derivable from these quantities). Unbiased parameter estimates can then be recovered from existing (biased) posterior chains by implementing an importance sampling scheme which accounts for this coordinate transformation, provided the chains are not too sparsely sampled in their low probability regions. Specifically, one can sample from a distribution $p_1(\vec{x})$ by reweighting samples from a different distribution $p_2(\vec{x})$. An example of this reweighting scheme as applied to a selection of DR25 targets is shown in Figure \ref{fig:b_posteriors_before_and_after}. A caveat is there is increased sampling error since $p_2(\vec{x})$ is a different distribution and the samples are not optimally distributed in $p_1(\vec{x})$. In essence there are smaller number of ``effective samples'' after reweighting. Care must therefore be taken to ensure that Jacobian-corrected posteriors are reliable, and the reweighting scheme we have outlined here should not be applied blindly.

\begin{figure}
    \centering
    \includegraphics[width=0.45\textwidth]{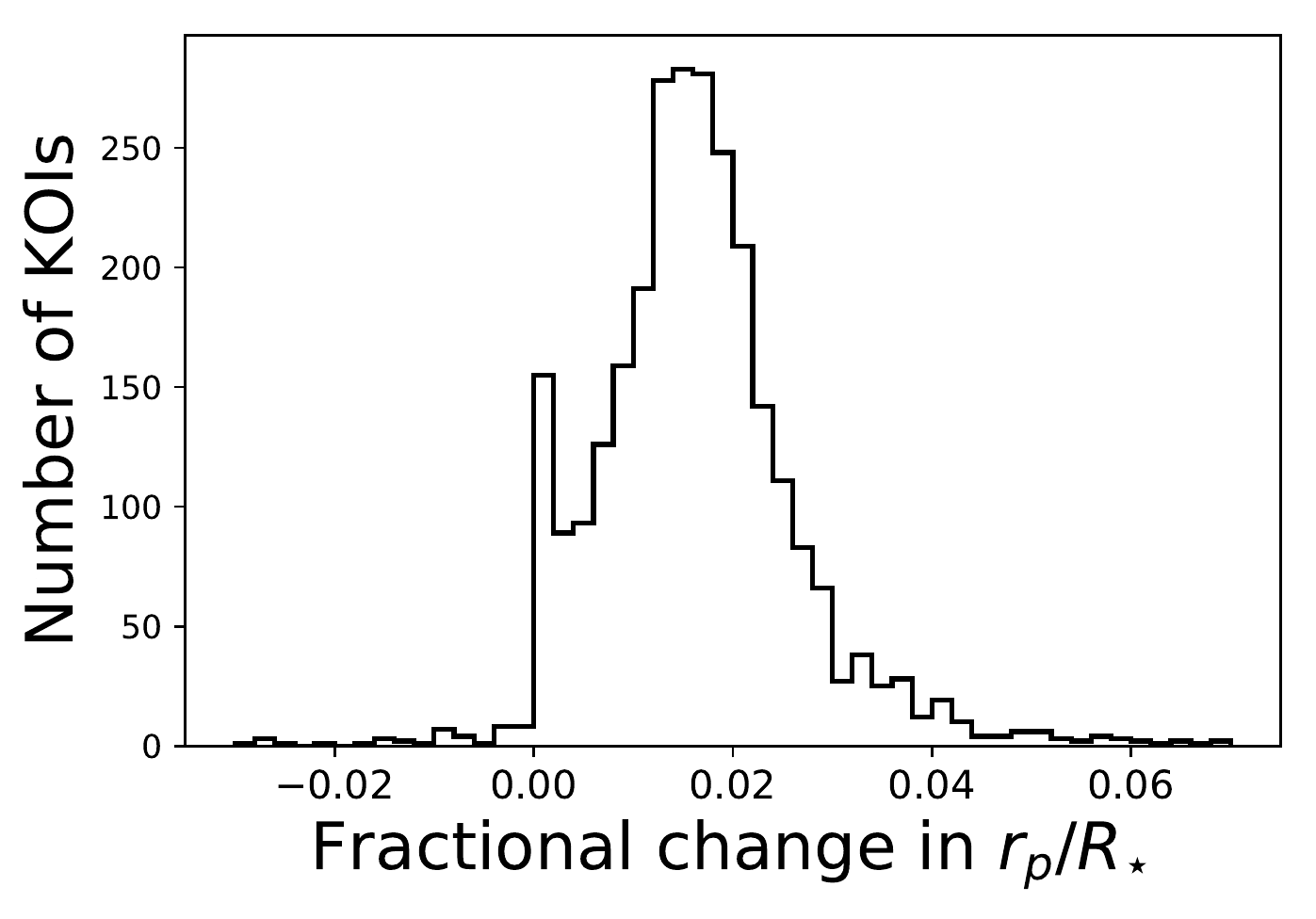}
    \caption{Fractional change in median planet-to-star radius ratio for all planet candidates after correcting posterior chains from DR25 \citep{Thompson2018} using the Jacobian reweighting scheme described in \S\ref{sec:kepler}. $5\sigma$ outliers have been iteratively clipped in order to eliminate spurious values that are expected to arise due to insufficient sampling of low probability regions. There is a spike at $\delta_r = 0$, indicating that some subset of targets were accurately measured, but the majority of targets are distributed around $\delta_r/r = 1.6\%$. }
    \label{fig:reweighted_delta_r}
\end{figure}

Because $b$ is covariant with $r$ (interacting via non-zero limb darkening), any bias on $b$ translates to a bias on $r$. For measurements in the final \Kepler data release, DR25, we find this covariance has produced a $1.6\%$ median systematic underestimate of $r$ (Figure \ref{fig:reweighted_delta_r}), extending as high as $\sim$6\% for some targets. This offset is comparable to the fractional uncertainty on $R_{\star}$ \citep{GaiaDR2, Berger2018} and so makes up a sizeable portion of the error budget for \Kepler planetary radii. While a few percent difference in planetary radius for a \textit{single} planet may be sub-significant, a systematic bias of a few percent on \textit{all} planetary radii will significantly impact our interpretation of population demographics -- for example, the precise characteristics of the radius valley \citep{Fulton2017} -- thereby altering our understanding of the processes by which planets form and evolve.

\section{Summary and conclusions}\label{sec:conclusion}

In this work, we explored the the biases that result from using the popular stellar pseudo-density, $\tilde{\rho}$, as a parameter in light curve fits. Adopting a linear prior on this parameter results in a biased distribution on impact parameter due to the Jacobian that arises from the non-linear relationship between $\tilde{\rho}$ and transit duration, $T_{14}$. Biased inferences on $b$ lead to biased inferences on $r$ due to covariances between the two parameters. We confirmed that the these biases are present in \Kepler modeling that used $\tilde{\rho}$ as a fitting parameter, and we presented a method for de-biasing the distributions. 

Although the $\tilde{\rho}$ bias may be resolved by using $\log\tilde{\rho}$ in place of $\tilde{\rho}$ (or, equivalently, placing log-uniform priors on $\tilde{\rho}$), we prefer sampling in duration over $\tilde{\rho}$ for aesthetic and conceptual reasons. To avoid inducing biases, we recommend sampling directly in duration $T_{14}$ or replacing $T_{14}$ with the true stellar density and orbital eccentricity vector, i.e. $\{\rho_{\star}, \sqrt{e}\sin\omega, \sqrt{e}\cos\omega\}$.

This work focused on the biases induced from using $\tilde{\rho}$ directly as a fitting parameter; similar biases may arise when using any related parameterization, for example $a/R_{\star}$, which is a popular choice \citep[e.g.][]{Crossfield2015, David2016, Stassun2017}. As with $\tilde{\rho}$, adopting a log-uniform prior rather than a linear prior on $a/R_{\star}$ avoids the unwanted bias. A log-uniform prior is a common choice, so most analyses which have used $a/R_{\star}$ as a fitting parameter are probably unaffected by the bias. However, one should always verify what priors were adopted when interpreting the results of any transit model.

\begin{acknowledgments}
We thank the anonymous referee for reviewing and providing comments which improved the quality of this manuscript. We are grateful to Jason Eastman, Dan Fabrycky, Dan Foreman-Mackey, Jason Rowe, Josh Winn, and Jon Zink for helpful conversations about this work.

G.J.G., M.G.M., and E.A.P. acknowledge support from NASA Astrophysics Data Analysis Program (ADAP) grant (80NSSC20K0457). E.A.P. acknowledges support from the Alfred P. Sloan Foundation. M.G.M acknowledges support from the UCLA Cota-Robles Graduate Fellowship. 

This study made use of data products from the \Kepler mission hosted on the \citealt{koidr25}. Some of the data were obtained from the Mikulski Archive for Space Telescopes (MAST) at the Space Telescope Science Institute. These data can be accessed via \dataset[10.26133/NEA5]{\doi{10.26133/NEA5}}. This study also made use of computational resources provided by the University of California, Los Angeles and the California Planet Search.

\end{acknowledgments}

\vspace{5mm}
\facilities{Kepler}

\software{astropy \citep{astropy:2018},
          exoplanet \citep{exoplanet:2021},
          numpy \citep{numpy:2020}, 
          PyMC \citep{pymc3:2016},
          scipy \citep{scipy:2020},
          starry \citep{starry:2019}
          }

\bibliography{pseudo_density}
\bibliographystyle{aasjournal}

\appendix
\section{Derivation of Jacobian for $T_{14} \rightarrow \tilde{\rho}$}

\noindent
In this section, we derive the Jacobian of the coordinate transformation $T_{14} \rightarrow \tilde{\rho}$. The pseudo-density derived by \citet{SeagerMallenOrnelas2003} is
\begin{equation}\label{app:rho_smo}
    \tilde{\rho} \equiv \Bigg(\frac{4\pi^2}{P^2G}\Bigg) 
    \Bigg(\frac{(1+r)^2 - b^2\big(1 - \sin^2[\pi T/P]\big)}{\sin^2[\pi T/P]}\Bigg)^{3/2}
\end{equation}
where all variables are defined as in previous sections. For notational clarity, we also define $T \equiv T_{14}$ and make the simplifying assumption $r \approx \sqrt{\Delta F}$, where $\Delta F$ is the fractional change in flux.
Substituting terms
\begin{equation}\label{app:xyz_sub}
\begin{aligned}
    &x = 4\pi^2/(P^2 G) \\
    &y = (1+r)^2 \\
    &z = \sin^2[\pi T/P]
\end{aligned}
\end{equation}
yields
\begin{equation}
    \tilde{\rho} = x \Bigg(\frac{y -b^2(1-z)}{z}\Bigg)^{3/2}.
\end{equation}
By the chain rule,
\begin{equation}\label{app:chain_rule}
    \frac{d\tilde{\rho}}{dT} =\frac{d\tilde{\rho}}{dz} \frac{dz}{dT}.
\end{equation}
The first term is
\begin{equation}\label{app:drho_dz}
    \frac{d\tilde{\rho}}{dz} = -\frac{3x}{2}\Bigg(\frac{y-b^2}{z^2}\Bigg)\Bigg(\frac{y - b^2(1-z)}{z}\Bigg)^{1/2}
\end{equation}
and the second term is
\begin{equation}\label{app:dz_dt}
    \frac{dz}{dT} = \frac{\pi}{P}\sin\Big[\frac{2\pi T}{P}\Big]
\end{equation}
Combining equations \ref{app:xyz_sub}, \ref{app:chain_rule}, \ref{app:drho_dz}, and \ref{app:dz_dt} yields the exact Jacobian
\begin{equation}\label{app:exact_jacobian}
     J = \frac{d\tilde{\rho}}{dT} = 
     -\frac{6\pi^3}{P^3G} 
     \Bigg(\frac{(1+r)^2 - b^2}{\sin^4[\pi T/P]}\Bigg) \\
     \Bigg(\frac{(1+r)^2 - b^2\big(1-\sin^2[\pi T/P]\big)}{\sin^2[\pi T/P]}\Bigg)^{1/2} 
     \sin\Big[\frac{2\pi T}{P}\Big]
\end{equation}
Making the small angle approximation $\sin\phi \approx \phi$ (assuming $\pi T \ll P$) and collecting terms yields  
\begin{equation}
    J = 
     -\frac{12\pi^3}{P^3G}
    \Bigg((1+r)^2 - b^2\Bigg)
    \Bigg((1+r)^2 - b^2\big(1  - [\pi T/ P]^2 \big)\Bigg)^{1/2}
    \Bigg(\frac{\pi T}{P}\Bigg)^{-4}.
\end{equation}
Once again taking advantage of $\pi T \ll P$ simplifies the expression further to
\begin{equation}\label{app:approx_jacobian}
    J = 
     -\frac{12\pi^3}{P^3G}
    \Bigg((1+r)^2 - b^2\Bigg)^{3/2}
    \Bigg(\frac{\pi T}{P}\Bigg)^{-4}.
\end{equation}

\break
\section{Derivation of Jacobian for $T_{14} \rightarrow \ln \tilde{\rho}$}

To derive the Jacobian of the transformation $T \rightarrow \ln \tilde{\rho}$, we note that

\begin{equation}\label{app:log_chain}
    \frac{d\ln\tilde{\rho}}{dT} = \frac{1}{\tilde{\rho}}
    \frac{d\tilde{\rho}}{dT}.
\end{equation}
Adopting our usual approximations $\sin\phi\approx\phi$, $\pi T \ll P$, we may rewrite Equation \ref{app:rho_smo} in the simplified form
\begin{equation}\label{app:rho_approx}
    \tilde{\rho} \equiv \Bigg(\frac{4\pi^2}{P^2G}\Bigg) 
    \Bigg((1+r)^2 - b^2\Bigg)^{3/2}
    \Bigg(\frac{\pi T}{P}\Bigg)^{-3}.
\end{equation}
Combining Equations \ref{app:approx_jacobian}, \ref{app:log_chain}, and \ref{app:rho_approx} and cancelling terms yields
\begin{equation}\label{app}
    \frac{d\ln\tilde{\rho}}{dT} =
    - \frac{3}{T}.
\end{equation}
We see that $d\ln\tilde{\rho}/dT$ is independent of $b$.

\section{Derivation of Jacobian for $T_{14} \rightarrow \lowercase{a}/R_{\star}$}

To derive the Jacobian of the transformation $T \rightarrow a/R_{\star}$, we define $\alpha \equiv a/R_{\star}$ and recognize that from \citet{SeagerMallenOrnelas2003} (their Equations 8 \& 9),
\begin{equation}\label{app:alpha}
    \alpha = \Bigg(\frac{4\pi^2}{P^2G}\Bigg)^{-1/3} \tilde{\rho}^{1/3}
\end{equation}.

By the chain rule,
\begin{equation}\label{app:alpha_rho}
    \frac{d\alpha}{dT} = \frac{d\alpha}{d\tilde{\rho}}\frac{d\tilde{\rho}}{dT}.
\end{equation}

The first term is
\begin{equation}\label{app:alpha_rho_chain}
    \frac{d\alpha}{d\tilde{\rho}} = \frac{1}{3}\Bigg(\frac{4\pi^2}{P^2G}\Bigg)^{-1/3} \tilde{\rho}^{-2/3}
\end{equation}

and the second term we derived previously. Adopting our usual approximations $\sin\phi\approx\phi$, $\pi T \ll P$ and combining Equations \ref{app:approx_jacobian}, \ref{app:rho_approx}, \ref{app:alpha_rho}, and \ref{app:alpha_rho_chain} yields
\begin{equation}\label{app:da_dT}
    \frac{d\alpha}{dT} = 
    -\frac{\pi}{P}
    \Bigg((1+r)^2 - b^2\Bigg)^{1/2}
    \Bigg(\frac{\pi T}{P}\Bigg)^{-2}.
\end{equation}

\section{Derivation of Jacobian for $T_{14} \rightarrow \ln \lowercase{a}/R_{\star}$}

To derive the Jacobian of the transformation $T \rightarrow \ln a/R_{\star}$, we note that
\begin{equation}\label{app:log_alpha}
    \frac{d\ln\alpha}{dT} = \frac{1}{\alpha}
    \frac{d\alpha}{dT}
\end{equation}
where as before $\alpha \equiv a/R_{\star}$. Following our usual strategies and combining Equations \ref{app:alpha}, \ref{app:da_dT}, and \ref{app:log_alpha}, we arrive at
\begin{equation}
    \frac{d\ln\alpha}{dT} = -\frac{1}{T}
\end{equation}
which is independent of $b$.
\

\end{document}